# Spicules: velocity, acceleration and scatter plots

Leonard A Freeman

**Abstract**

Spicules and other solar jets such as bright points and fibrils generally show a parabolic height-time relationship, which means that each spicule has a constant deceleration. However the deceleration is only constant for a particular spicule and varies widely from one spicule or jet to another. Nonetheless the careful observations of a number of researchers show that the distance - time relationship is parabolic to a high level of precision. The measurements for heights, maximum velocities, decelerations and flight times are normally presented as histograms or scatter plots, which allow some general trends to be observed. The published results show a clear correlation between the maximum velocity and the deceleration of spicules on scatter plots. This correlation has been claimed to show a linear relation between the acceleration and the maximum velocity of a jet. This linear relationship has been used to help model the mechanisms responsible for the jets.

However it is proposed here that the relation between velocity and acceleration is given by the normal equations of motion for constant acceleration and consequently the relationship is non-linear. Other correlations are also examined and the implications for spicule mechanisms are considered.

**Introduction**

Spicules are transient, hair-like jets that are common all over the surface of the sun. They typically reach heights of 1- 20 km and have lifetimes from 1 to 20 minutes or more. Recent measurements such as those made using the Swedish Solar Telescope or from spacecraft such as Hinode provide greater spatial and temporal resolution, allowing more accurate observations to be made of the kinematics of spicules, see Hansteen et al (2006), De Pontieu et al (2007a, 2007b), Anan et al (2010), Zhang et al (2012), Langangen et al (2008), Pereira et al (2012), van der Voort et al (2013) and more recently Loboda et al (2017).

These authors find a good parabolic (x,t) fit to the motion of spicules and jets, where x is the height of the spicule (or more accurately the length along its path, as spicules are usually inclined) and t is the time from its starting point. A typical graph of a 2000 km height spicule with a lifetime of 200 seconds is shown in Figure 1.

Figure 1

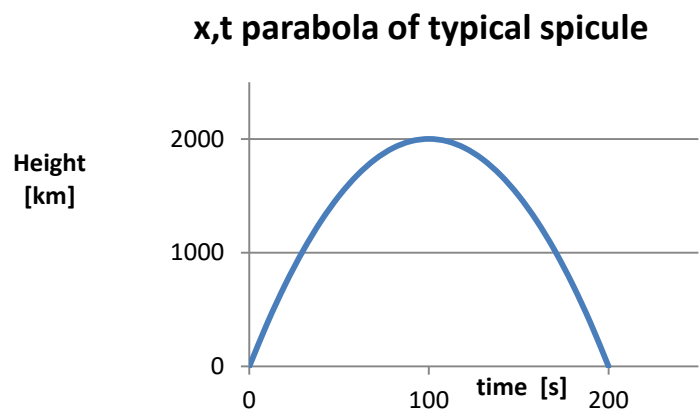

x,t parabola of typical spicule

The (x,t) curve allows the four key parameters Vmax, acceleration, height and lifetime to be calculated.

Although we will often refer only to spicules, this will be taken to include similar types of jet, such as bright points, mottles or dynamic fibrils that also have constant acceleration.

A spicule has a fast impulsive start with a maximum velocity, its launch velocity. It then rises up, usually along a path inclined to the vertical. The path is considered to be a local magnetic field line. Spicules closer to the sun's poles are closer to the vertical.

After the spicule's rapid launch, it decelerates at a constant rate, reaches its maximum height,when its velocity is zero. It then accelerates back down at the same rate, so that its final velocity is the same as its start velocity. This is comparable to ballistic behaviour, but the difference is that the value of deceleration/acceleration is not the same as that due to solar gravity: it can be much lower or higher than solar gravity. But a consequence of the parabolic nature is that the deceleration is constant for any particular spicule. Remarkably, solar gravity appears to have little effect on the way jets rise and fall. The term "ballistic" is not used here, to avoid possible misunderstanding.

The results of the observations of spicule heights, lifetimes velocities and accelerations are usually presented as histograms, which show for example the frequency distribution of spicule heights or as a scatter plot of two variables such as (a,V) - acceleration against maximum velocity

In these results the scatter plot of (a,V) shows a clear positive correlation, but the points are widely scattered, making a definitive relationship difficult to establish.

Figure 2, for example shows an envelope for the many hundreds of results given by Langangen.et al.(2008).

Figure 2

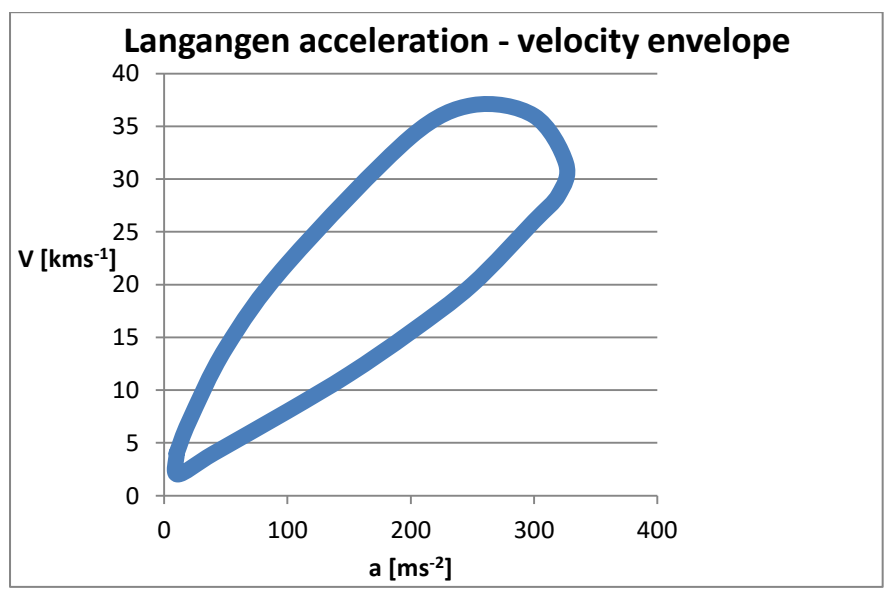

The scatter plot of Langangen et al (2008) indicates the large degree of scatter in (a,V) plots. The envelope includes many hundreds of data points, including the work of Hansteen et al (2006). Similar amounts of scatter are found by others.

Table 1, shows for example the wide variation in the acceleration, a, for spicules with a launch velocity of 20 km s$^{-1}$. The data is taken from the scatter plots.

| Scatter plot variation in a (ms$^{-2}$) for spicules or jets with Vmax = 20km s$^{-1}$ | | | | | | |
| --- | --- | --- | --- | --- | --- | --- |
| 80 - 240 | 70 - 290 | 90 - 240 | 75 - 380 | 70 - 270 | 50 - 400 | 100 - 300 |
| De Pontieu et al (2007a) | De Pontieu et al (2007b) | Langangen et al (2008) | Pereira et al (2012) | Van der Voort et al (2013) | Zhang et al (2012) | Anan et al (2010) |

Despite the large variation, scatter plots such as figure 2 show a clear correlation between the variables, but is it a linear relationship? Hansteen et al (2006), De Pontieu et al (2007b) and Tsiropoula et al (2012) consider that this shows a linear relationship between V and a. However the actual equation does not seem to be generally agreed. Loboda et al (2017) for example, looking for a linear relationship in their study of macrospicules, find the equation:

$$V = 340a + 51.3 \qquad (1)$$

Where V is in km s$^{-1}$ and a is in km s$^{-2}$

But they comment that the numbers in the equation differ by a factor of three compared with those found in other papers. It is important to note that Loboda et al (2017) were measuring macrospicules, which are much longer than most ordinary spicules – we see later that this is the cause of the apparent disagreement.

**Analysis and theory**

The parabolic behaviour of a range of jets has been established by many observations and is a consequence of constant acceleration for each individual jet.

If acceleration a = constant, then it is easy to show that this leads to the equations describing parabolic behaviour:

$$V = at \qquad (2)$$
$$V^2 = 2ah \qquad (3)$$
$$h = \frac{1}{2}at^2 \qquad (4)$$
$$h = \frac{1}{2}Vt \qquad (5)$$

Where h is the maximum height of a spicule. Its velocity is a maximum, V at the beginning and end of its flight, and zero at maximum height. Time t, is the time taken to rise from the start to its maximum height, or to fall back to its start position. Its lifetime is then twice this value, 2t.

**Explaining the scatter**

Producing an a,V graph is always going to produce a lot of scatter, because there is no single one-to-one relationship between V and a. As can be seen from equations (2) and (3), for example, velocity is a function of two variables, only one of which is acceleration. Equation (3) for example shows that V depends not only on a, but also the height, h.

To illustrate this, consider first a study of spicules of the same height, excluding all the others. Let us choose a height of some of the smaller spicules – say 1000 km. We then get a simple curve described, from equation (3) by:

$$V = \sqrt{2 \times 1000} \sqrt{a} \qquad (6)$$

All spicules of this height which have a parabolic nature will obey this equation.

Figure 3 shows this curve.

Figure 3

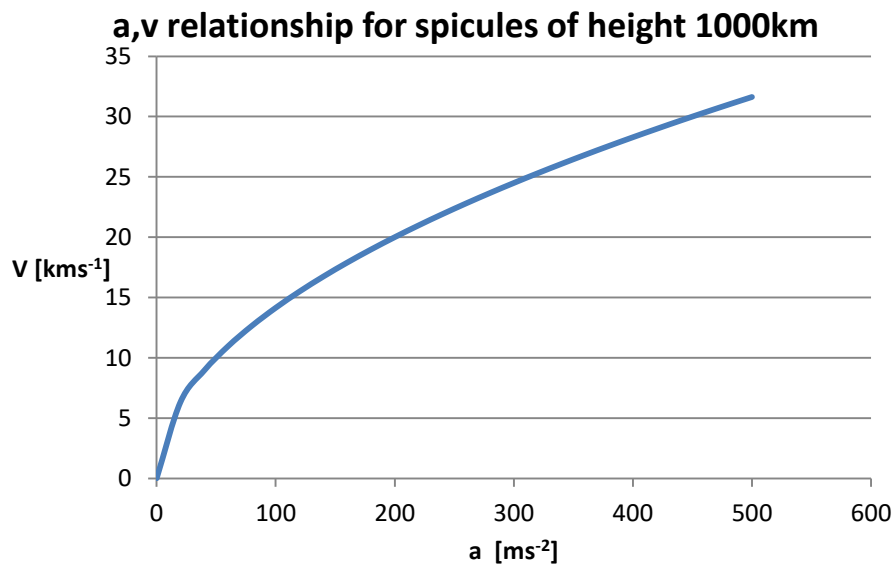

a,v relationship for spicules of height 1000km

The curve shown in figure 3 is simply a consequence of the parabolic (x,t) relationship, which is the same as that saying each individual spicule has a constant deceleration/acceleration.

The figure clearly shows that for all spicules of this height, the deceleration increases as the launch velocity of the spicule increases, and is given by equation (6). There is no scatter if only spicules of the same height are involved, apart from normal measurement errors.

Next we consider the curve for spicules that are much longer, say five times this height This produces a similar, but higher curve as seen in figure 4.

Figure 4

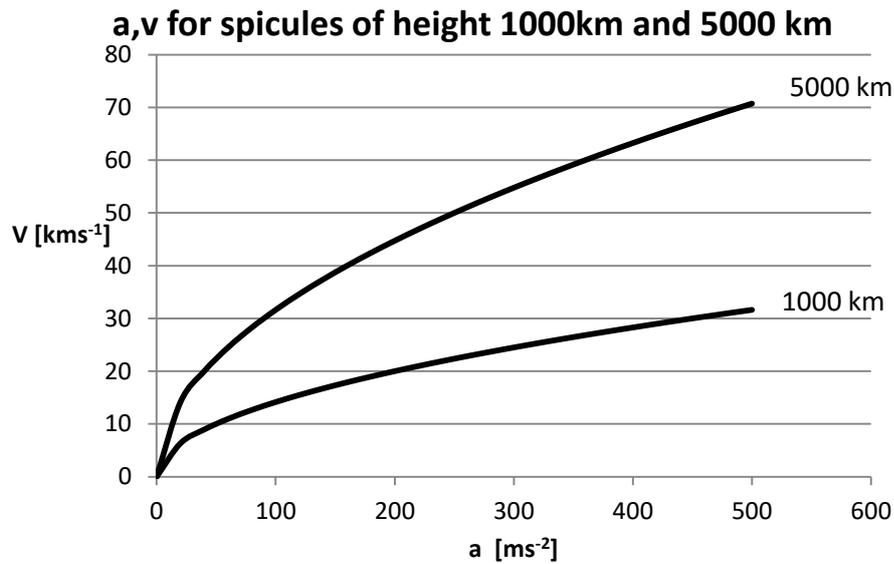

If we now choose a set of 100, say, random heights, between 1000 km and 5000 km, then instead of two fixed heights we have the scatter plot shown in figure 5, which is similar to the scatter graphs referred to above.

Figure 5

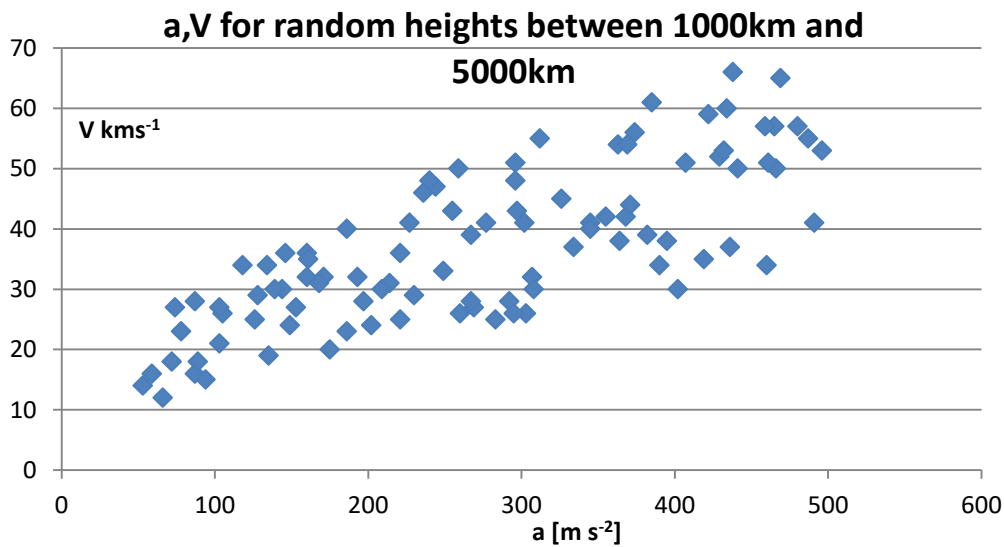

This shows that a random set of 100 data points from jets whose height lies between these two values shown will give a scatter plot of points lying between the two curves, similar to those reported. At first sight it appears that these points could have a linear correlation, but from equation (3) we can simply say that:

For all spicules of the same height, V is proportional to the square root of the acceleration, that is
$$V = k\sqrt{a} \qquad (7)$$

where k is a constant equal to √(2h). The scatter is not due to any measurement error, but mainly because the effect of the third variable, height in this case, has been omitted.

**Comparison with results**

Loboda et al (2017) have recently reported some interesting results on macrospicules. These rise to much greater heights than the average spicule and are the less common type and often found in coronal holes. They are also convenient here, since they have a more restricted range of heights, relatively speaking.

They provide measurements detail for 15 macrospicules, and figure 6 shows their data points marked with crosses

Figure 6

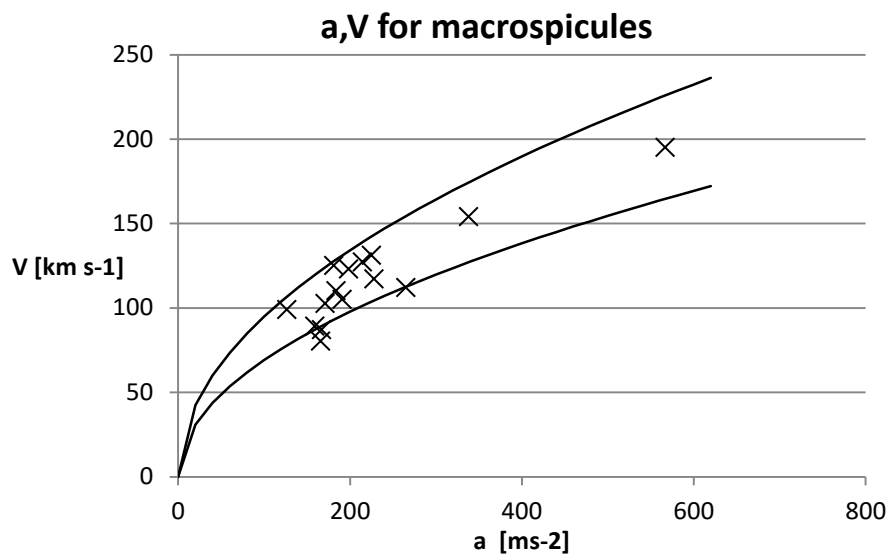

The graph also shows the curves corresponding to the lowest (23,900 km) and highest (45,000 km) heights of spicule they measured. Their data lies reasonably well between the two curves, which indicates again that the a,V relationship is not linear. We can now see the reason for the high values they obtained for their linear relation in equation (1). Their results are for much higher points on the V,a curve so the intercept will be much higher. Also their range of heights is more restricted, leading to a higher gradient.

Instead of using a scatter plot, we can bring a variety of results together using equation (3). If we plot V against √(2ah) we should get a straight line with gradient of one.

Figure 7 shows this plot, not only for the 15 points of Loboda et al, but also for 2 lower points measured from "standard" spicules. The lowest point on the curve is taken from the data of Langangen et al (2008), and is itself the mean of 230 measurements of dynamic fibrils. They give the mean values of time, acceleration and velocity respectively as 258 s, 18.6 kms$^{-1}$ and 142 ms$^{-2}$; the height is calculated from these values. The other point just above that is from 125 parabolic spicules measured by Anan (2010) et al, who report respectively mean values of height, acceleration and velocity as 1300 km, -510 ms$^{-2}$ and 34.4 kms$^{-1}$.

Figure 7

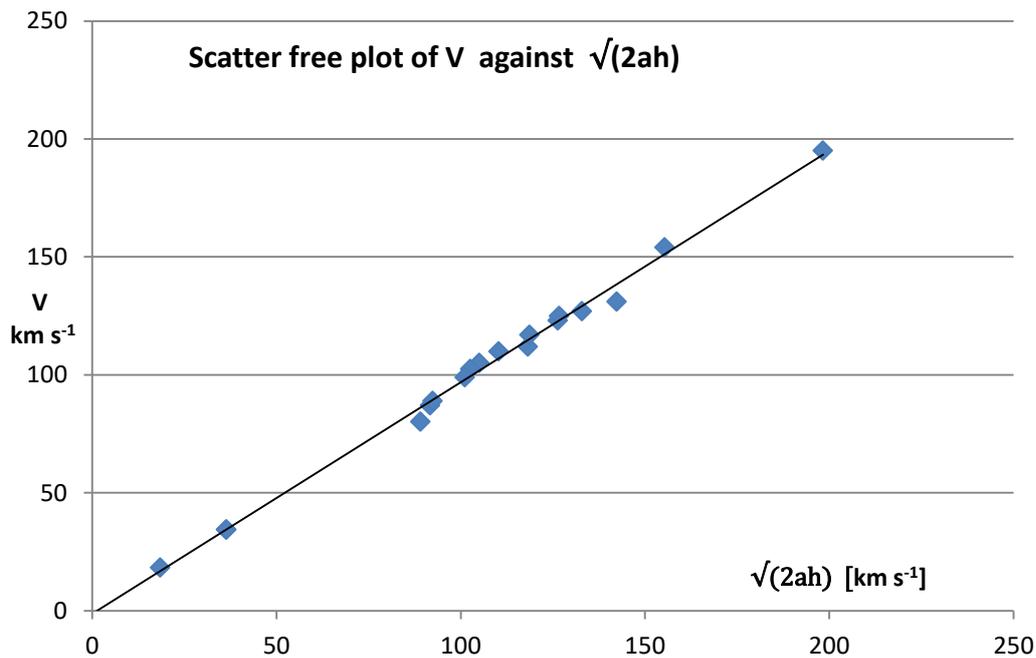

This very good agreement is only to be expected for objects with a constant deceleration. It demonstrates that the large scatter seen in a,V plots disappears when the third variable, height is included

However even this graph simply reflects the constant acceleration property, but provides no further information on the actual mechanism.

In contrast to scattergrams, the role of histograms showing the distribution of heights, velocities and lifetimes is essential in the understanding of spicules. Any model must also account for the range of heights and lifetimes.

**Other observed correlations**

Scatter plots of other variables can also show correlations. For example, De Pontieu et al (2007a) report an anticorrelation between duration, t and acceleration, a. All such correlations are simply a reflection of one of the four equations of motion, together with the fact that the range of heights or velocities is restricted. To examine this (t,a) correlation we will use 100 random heights, chosen as for figure 5, and use their (a,V) data to find the values of t, from equation (2). This gives for the individual points $t = v/a$, but as t is only half the total time we must double this, to get $t' = 2v/a$. The graph for the results of t' against a is shown in figure 8. The restriction on height values automatically produces a corresponding restriction on velocity values.

Figure 8

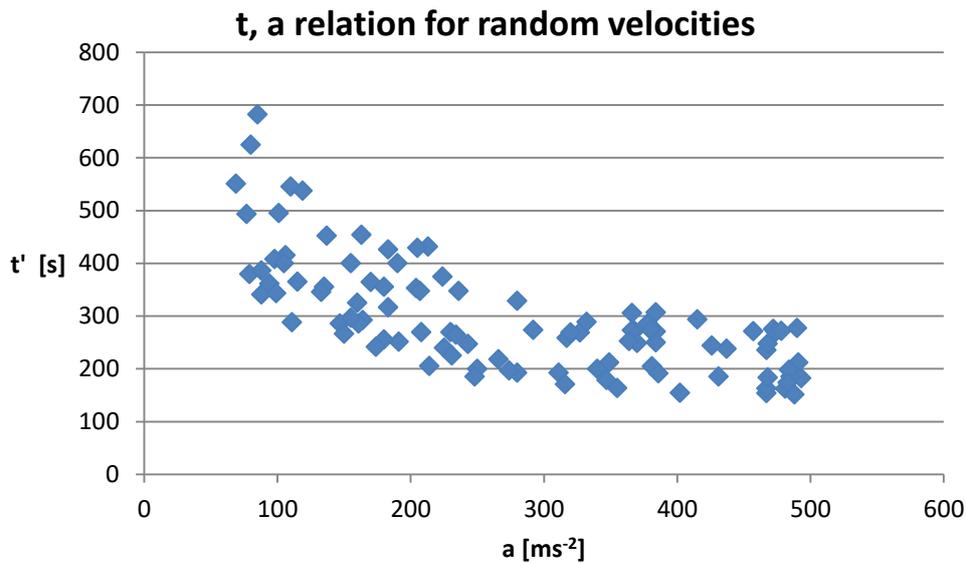

This curve is reflecting the equation t = V/a , together with the fact that the values of V (and h) are restricted. If V could have any value, the graph would be covered with points and the curve would not be discernible.

Again, we have a graph that seems to be conveying important information, but closer analysis now reveals that it is just a consequence of one of the equations of motion, t=V/a for objects with constant acceleration. The "scatter" is due solely to the omission of the variable, V.

**Should the third variable be height, h, or time, t ?**

When examining the a,V relationship it can be argued that we should select time, t, as the third variable instead of height, h. This is an equally valid alternative, given by equation (2). If the analysis is repeated, using t instead of h, we come to the same basic conclusions. The a,V scatter graph can now be considered as a set of lines of the form V = $t_n$ a . Each line then represents all spicules with the same lifetime, $t_n$.

Figure 9 illustrates this with two lines, the lower line being spicules with the shortest life and the upper line representing those with the longest life. In between these will be the actual cloud of random data points, so again there is no linear relationship between a and V, except for spicules with the same lifetime. Note that as before, the scatter of points increases as either a or V increase, or indeed the range of lifetimes. Arbitrary units are used for this graph.

Figure 9

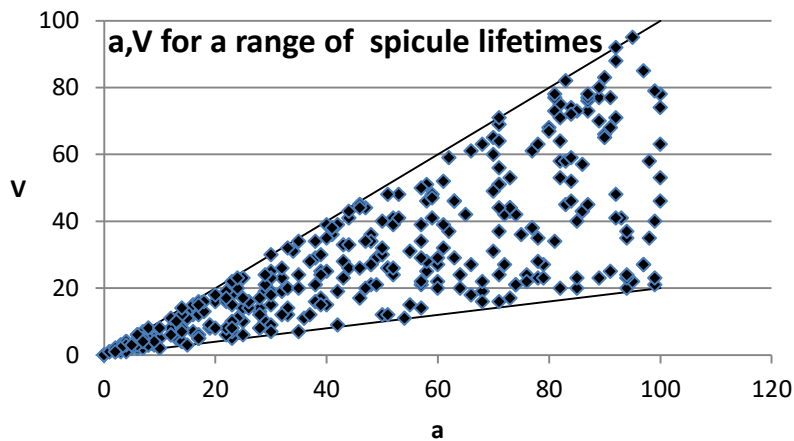

**Conclusion and discussion**

This analysis shows that it is unhelpful to look for a relationship between acceleration and velocity, if the spicule is seen to have a parabolic (x,t) form. That (a,V) connection is fixed by the equations of motion, which means that there is no extra relationship involving just the two quantities, because they depend on a third variable, either height or time. The missing quantity, such as height, is usually known or can be calculated. When this "missing" variable is included, the scatter disappears.

This in turn means that simulations which produce a linear relationship between (a,V) are likely to be incorrect, as they conflict with the laws of motion for constant acceleration. However, if the simulation clearly produces parabolic behaviour, then it has met one of the key tests in describing spicules. But the examination of the relationship between (a,V) alone will produce little extra information. Histograms of the variables, however are vital in establishing the range of heights, velocities, accelerations and lifetimes.

So what does this mean for models of the mechanism which control spicule behaviour? Clearly the first necessary condition is that any model must produce a constant deceleration/acceleration, which is equivalent to saying that the motion is parabolic in its height-time motion.

The a,V relationship is then simply a reflection of the parabolic behaviour and is given in the equations (2) to (5).

The other requirement is that the model must provide an explanation for the typical heights (or equivalently, lifetimes) of spicules. If these two conditions are satisfied, we have a suitable model.

In summary, suitable models must give

1. A constant deceleration
2. An explanation for the heights of spicules

Also required, of course, is a launch mechanism to inject the spicules into the chromosphere, but here that is considered a separate problem.

Various mechanisms have been examined to explain spicules and currently the application of magnetoacoustic shocks to the chromosphere looks a promising candidate.

But is it possible that local magnetic fields could be responsible for the parabolic behaviour? It is generally recognised that magnetic fields control much of the behaviour of solar jets. For example their inclination is controlled by the direction of the local magnetic field and the height of spicules varies from one region of the sun to another, possibly dependent on the local field strength. Could the magnetic field also account for the parabolic behaviour?

Freeman (2013) has suggested that magnetised plasma may be attracted to the source of a magnetic field, and that the force of attraction is related to the field gradient. If this is correct, it would provide a suitable explanation of spicule behaviour, in attracting solar jets back to the surface.

The particular mechanism proposed in that paper has not been confirmed, but if it, or a similar mechanism can be validated, it would give a simple magnetic explanation for the parabolic behaviour of spicules.

Leonard Freeman
23 Hope Street
Cambridge CB1 3NA
UK
leonard.freeman@ntlworld.com